\documentclass[RNAAS,twocolumn,tighten]{aastex631}
\usepackage{graphicx}
\usepackage[suffix=]{epstopdf}
\usepackage{amsmath}
\usepackage{url}
\usepackage{xspace}
\usepackage{mathrsfs}

\begin{document}

\title{Real-Time Technosignature Strategies with SN\,2023ixf}


\author[0000-0002-0637-835X]{James R. A. Davenport}
\affiliation{Astronomy Department, University of Washington, Box 951580, Seattle, WA 98195, USA}

\author[0000-0001-7057-4999]{Sofia Z. Sheikh}
\affiliation{SETI Institute, 339 N Bernardo Ave Suite 200, Mountain View, CA 94043, USA}
\affiliation{Berkeley SETI Research Center, University of California, Berkeley, CA 94720, USA}

\author[0000-0002-0161-7243]{Wael Farah}
\affiliation{SETI Institute, 339 N Bernardo Ave Suite 200, Mountain View, CA 94043, USA}
\affiliation{Berkeley SETI Research Center, University of California, Berkeley, CA 94720, USA}

\author[0000-0002-5956-5167]{Andy Nilipour}
\affiliation{Department of Astronomy, Yale University, Steinbach Hall, 52 Hillhouse Ave, New Haven, CT 06511}

\author[0009-0009-1019-3890]{B\'arbara Cabrales}
\affiliation{Department of Astronomy, Smith College, Northampton, MA 01063, USA}
\affiliation{Berkeley SETI Research Center, University of California, Berkeley, CA 94720, USA}

\author[0000-0003-4823-129X]{Steve Croft}
\affiliation{SETI Institute, 339 N Bernardo Ave Suite 200, Mountain View, CA 94043, USA}
\affiliation{Berkeley SETI Research Center, University of California, Berkeley, CA 94720, USA}

\author[0000-0002-3430-7671]{Alexander W. Pollak}
\affiliation{SETI Institute, 339 N Bernardo Ave Suite 200, Mountain View, CA 94043, USA}
\affiliation{Berkeley SETI Research Center, University of California, Berkeley, CA 94720, USA}

\author[0000-0003-2828-7720]{Andrew P. V. Siemion}
\affiliation{SETI Institute, 339 N Bernardo Ave Suite 200, Mountain View, CA 94043, USA}
\affiliation{Berkeley SETI Research Center, University of California, Berkeley, CA 94720, USA}
\affiliation{Department of Physics and Astronomy, University of Manchester, UK}
\affiliation{University of Malta, Institute of Space Sciences and Astronomy, Msida, MSD2080, Malta}

\begin{abstract}
Several technosignature techniques focus on historic events such as SN\,1987A as the basis to search for coordinated signal broadcasts from extraterrestrial agents. The recently discovered SN\,2023ixf in the spiral galaxy M101 is the nearest Type II supernova in over a decade, and will serve as an important benchmark event. Here we review the potential for SN\,2023ixf to advance ongoing techonsignature searches, particularly signal-synchronization techniques such as the ``SETI Ellipsoid''. We find that more than 100 stars within 100\,pc are already close to intersecting this SETI Ellipsoid, providing numerous targets for real-time monitoring within ${\sim}3^\circ$ of SN\,2023ixf. We are commencing a radio technosignature monitoring campaign of these targets with the Allen Telescope Array and the Green Bank Telescope.
\end{abstract}

\section{Introduction}

Signal synchronization has long been discussed as a useful strategy for facilitating interstellar communication. In this framework, extraterrestrial agents would use noteworthy events such as galactic novae \citep{makovetskii1977} or supernovae \citep{lemarchand1994} to coordinate signal transmission. This approach provides an elegant solution for when and where to search for tecnhosignatures \citep{tarter2001,wright2018c}. 

\citet{davenport2022} recently reviewed the ``SETI Ellipsoid'' approach, noting that Gaia EDR3 parallaxes \citep{gaia_edr3} enable precise predictions for when coordinated transmissions from nearby stars observing e.g. SN\,1987A would arrive. These approaches have been used recently to explore coordination with historic astronomical events (Nilipour et al. {\it in press}), and a range of sky surveys (Cabrales et al. {\it in prep}). They are especially useful for coordinating observations around new events being actively observed by a wide range of facilities.

\section{The SETI Ellipsoid for SN\,2023ixf}

SN\,2023ixf was discovered near the spiral galaxy M101 on 2023 May 19 by Mr. K. Itagaki\footnote{\url{https://www.wis-tns.org/object/2023ixf}}, with the earliest detection from ZTF on 2023 May 17 \citep{bellm2019}.
At ${\sim}6.19$\,Mpc \citep{mager2013}, this is the closest Type II SNe in a decade, and is already generating excitement as a new benchmark core collapse event. As such, SN\,2023ixf offers a prime opportunity for real-time searches for coordinated technosignature signals. 

\begin{figure*}[!ht]
\centering
\includegraphics[height=3.2in]{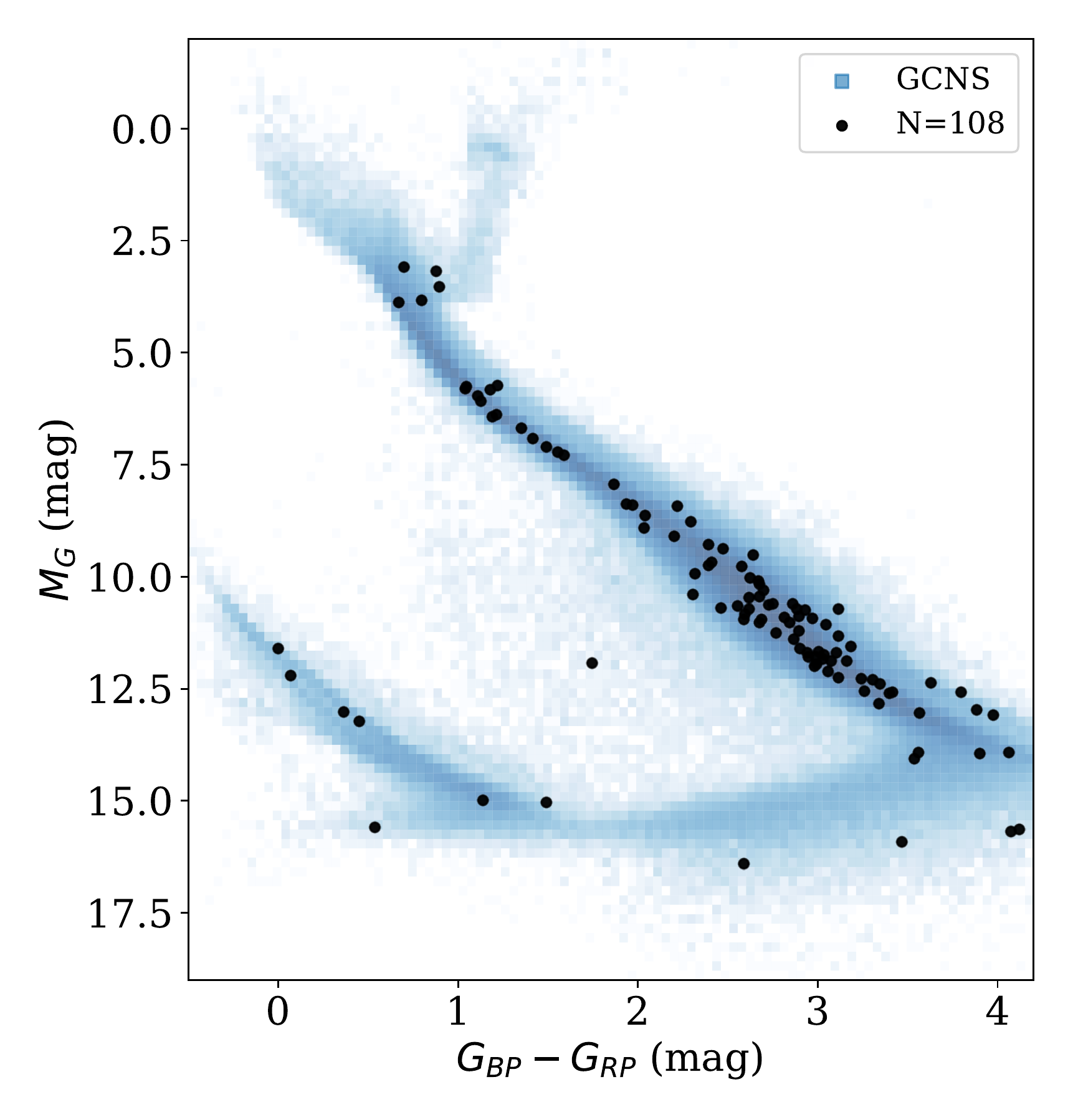}
\includegraphics[height=3.2in]{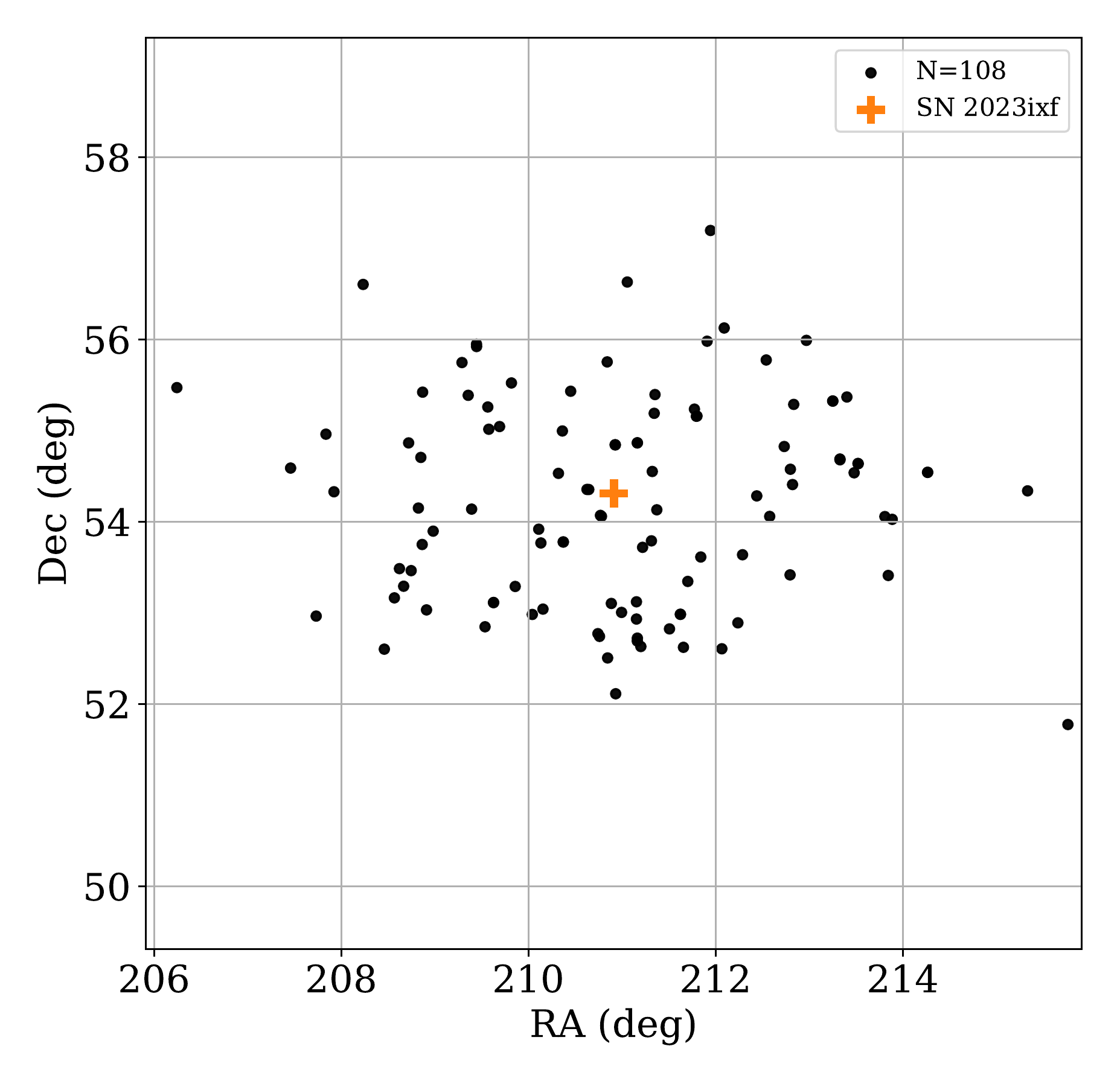}
\caption{
Left: Color--magnitude diagram for 108 stars  within 0.1\,ly of the SETI Ellipsoid defined by SN\,2023ixf as of MJD 60094 (black points), with the 100-pc Gaia sample for reference (blue contours). These Ellipsoid-crossing targets already span a range of stellar types. 
Right: Sky positions for the Ellipsoid-crossing stars (black points), which lay in a ${\sim}3^\circ$ region around SN\,2023ixf (orange cross).
}
\label{fig:elip}
\end{figure*}

Following \citet{davenport2022}, we have searched the Gaia 100\,pc star sample \citep{gaia-collaboration2021} for targets that are within 0.1 ly of the SETI Ellipsoid defined by SN\,2023ixf, which could plausibly broadcast a signal coordinated with the supernovae. 
SN\,2023ixf was discovered during its rise phase, and Type II SNe typically remain bright in the optical for several weeks. We have opted to use the discovery date of 2023 May 19 for our calculations. Our choice of a 0.1\,ly tolerance is motivated by both parallax uncertainties from Gaia, and the natural timescale of the SNe event. 
We find there are (as of MJD 60094) 108 stars within 0.1\,ly of this Ellipsoid, shown in Figure \ref{fig:elip}. As the light from SN\,2023ixf propagates, ${\sim}15$ stars per week enter this 0.1\,ly sample, and the projected size of the Ellipsoid grows by ${\sim}2^\circ$ per year.

\section{Real-Time Monitoring}

SN\,2023ixf provides a unique opportunity to monitor the SETI Ellipsoid of a nearby supernova within weeks of its discovery. The Allen Telescope Array (ATA) is an ideal instrument for follow-up because it was designed specifically for radio technosignature searches: it has a wide instantaneous bandwidth, a large field of view, high spectral resolution beamforming capabilities from its log-periodic feeds, a design with a large number of small dishes \citep{welch2009allen}, and a revamped state-of-the-art digital signal processing system. During the recent refurbishment of the array, dedicated time has been set aside for target-of-opportunity studies for technosignature applications \citep{Perez2022_RNAAS}, in addition to other astronomical follow-up \citep[e.g.][]{bright2023precise}.

We performed preliminary radio technosignature observations for 97 of the Ellipsoid-crossing main sequence stars with a two-beam survey using 20 6.1-m dishes, covering 1.4\,GHz of bandwidth centered at 4.9\,GHz. These observations produced Stokes-I filterbank files with\,1-Hz spectral resolution and 15-s time resolution, which will be searched with the \texttt{turboSETI} Doppler-drifting narrow-band technosignature search code  \citep{enriquez2019turboseti}.

Additionally, on 2023 May 31 we performed observations of the nearest (14.5\,pc) Ellipsoid-crossing star in the sample, HIP\,70497, using the L-band receiver at the Robert C.\ Byrd Green Bank Telescope; a second star in the sample, LHS\,6266, also fell within the GBT primary beam. Data were acquired using the Breakthrough Listen backend \citep{macmahon2018,lebofsky2019} using Listen's standard observing mode, an ABACAD cadence with 5 minutes per position, following the procedure outlined by \citet{franz2022} and others. Data will be processed with \texttt{turboSETI}.

We intend to revisit the Ellipsoid once a month for the next few months as new stars enter the sample, and are open to synchronizing our observations with other multiwavelength facilities.

\vspace{0.1in}
JRAD acknowledges support from the DiRAC Institute in the Department of Astronomy at the University of Washington. The DiRAC Institute is supported through generous gifts from the Charles and Lisa Simonyi Fund for Arts and Sciences, and the Washington Research Foundation. 
The Allen Telescope Array (ATA) refurbishment program and its ongoing operations have received substantial support from Franklin Antonio. Additional contributions from Frank Levinson, Greg Papadopoulos, the Breakthrough Listen Initiative and other private donors have been instrumental in the renewal of the ATA. Breakthrough Listen is managed by the Breakthrough Initiatives, sponsored by the Breakthrough Prize Foundation.  The Paul G. Allen Family Foundation provided major support for the design and construction of the ATA, alongside contributions from Nathan Myhrvold, Xilinx Corporation, Sun Microsystems, and other private donors. The ATA has also been supported by contributions from the US Naval Observatory and the US National Science Foundation. S.Z.S. acknowledges that this material is based upon work supported by the National Science Foundation MPS-Ascend Postdoctoral Research Fellowship under Grant No. 2138147. AN was funded as a participant in the Berkeley SETI Research Center Research Experience for Undergraduates Site, supported by the National Science Foundation under Grant No.~1950897.

\bibliographystyle{aasjournal}

\end{document}